\documentclass[%
 aip,
 jmp,%
 amsmath,amssymb,
 reprint,%
]{revtex4-1}
\usepackage{longtable}
\usepackage{graphicx}% Include figure files
\usepackage{dcolumn}% Align table columns on decimal point
\usepackage{bm}% bold math
\usepackage{amsmath}
\usepackage{color}
\usepackage{amssymb}
\usepackage{hyperref}% add hypertext capabilities
\begin{document}

\preprint{APS/123-QED}

\title{Two Dimensional Ferromagnetic Semiconductor: Monolayer CrGeS$_3$}

\author{Yulu Ren}
\email{ylren@stumail.ysu.edu.cn}
\affiliation{State Key Laboratory of Metastable Materials Science and Technology \& Key Laboratory for Microstructural Material Physics of Hebei Province, School of Science, Yanshan University, Qinhuangdao 066004, China}

\author{Yanfeng Ge}
\email{yfge@ysu.edu.cn}
\affiliation{State Key Laboratory of Metastable Materials Science and Technology \& Key Laboratory for Microstructural Material Physics of Hebei Province, School of Science, Yanshan University, Qinhuangdao 066004, China}

\author{Wenhui Wan}
\affiliation{State Key Laboratory of Metastable Materials Science and Technology \& Key Laboratory for Microstructural Material Physics of Hebei Province, School of Science, Yanshan University, Qinhuangdao 066004, China}

\author{Qiaoqiao Li}
\affiliation{State Key Laboratory of Metastable Materials Science and Technology \& Key Laboratory for Microstructural Material Physics of Hebei Province, School of Science, Yanshan University, Qinhuangdao 066004, China}

\author{Yong Liu}
\email{yongliu@ysu.edu.cn, ycliu@ysu.edu.cn}
\affiliation{State Key Laboratory of Metastable Materials Science and Technology \& Key Laboratory for Microstructural Material Physics of Hebei Province, School of Science, Yanshan University, Qinhuangdao 066004, China}

\date{\today}
\begin{abstract}
Recently, two-dimensional ferromagnetic semiconductors have been an important class of materials for many potential applications in spintronic devices. Based on density functional theory, we systematically explore the magnetic and electronic properties of CrGeS$_3$ with the monolayer structures. The comparison of total energy between different magnetic states ensures the ferromagnetic ground state of monolayer CrGeS$_3$. It is also shown that ferromagnetic and semiconducting properties are exhibited in monolayer CrGeS$_3$ with the magnetic moment of 3 $\mu_{B}$ for each Cr atom, donated mainly by the intense $dp$$\sigma$-hybridization of Cr $e_g$-S $p$. There are the bandgap of 0.70 eV of spin-up state in the monolayer structure when 0.77 eV in spin-down state. The global gap is 0.34 eV (2.21 eV by using HSE06 functional), which originates from bonding $dp\sigma$ hybridized states of Cr $e_g$-S $p$ and unoccupied Cr $t_{2g}$-Ge $p$ hybridization. Besides, we estimate that the monolayer CrGeS$_3$ possesses the Curie temperature of 161 K by mean-field theory.
\end{abstract}
\maketitle
%\begin{multicols}{2}
\section{introduction}
Graphene with honeycomb structure has been synthesized as extremely thin layers by mechanical exfoliation because of the weak van der Waals out-of-plane interaction, also opens a new field for the two-dimensional crystal with the new physics.\cite{1, 2, 3, 4, 5} Two-dimensional materials attracted intensive attention partially for their exotic properties. For example, 2D hexagonal boron nitride with a large bandgap of 4.64 eV is reported and possesses enormous potential in electronic and composite applications.\cite{6, 7, 8, 9, add7, add8} And Transition metal dichalcogenides with common structural formula MX$_2$ (M=Mo, W, Ti, etc. And X=S, Se, Te) are found to show electronic properties varying from metals to wide-gap semiconductors with interesting physical characteristics.\cite{10, 11, 12, 13, 14} MoS$_2$ possesses a transformation from indirect bandgap of bulk to direct bandgap of monolayer, leading to a promising luminescence quantum efficiency.\cite{10} Besides, many other compositionally diverse 2D materials have been predicted and in some cases synthesized.\cite{15, 16, 17, 18, 19}

The intrinsic ferromagnetic (FM) semiconductors possess FM and semiconducting properties simultaneously with the fully spin polarization. Compared with diluted magnetic semiconductors, FM semiconductors exhibit tremendous advantages indeed in carrier injection, detection, sensors, magnetic storage and emergent heterostructure devices.\cite{add6} However, a few FM semiconductors have been reported up to now limited by the difficult synthesis in experiment. And Mermin-Wagner theory illustrates the absence of intrinsic magnetization with 2D isotropic Heisenberg model at a finite temperature.\cite{20} Yet recent studies have shown 2D magnetic crystals can also exhibit magnetism with an Ising model. For example, layer-dependent FM van der Waals crystal like Fe$_3$GeTe$_2$ with an out-of-plane magnetocrystalline anisotropy process a controllable $T_C$ by an ionic gate,\cite{add4} and monolayer triiodide chromium (CrI$_3$) is Ising FM material with out-of-plane spin orientation and the $T_C$ of 45 K.\cite{add5} Also, nanosheets CrSiTe$_3$ with the indirect bandgap of 0.57 eV have been exfoliated successfully by Lin et al.\cite{21} In addition, intrinsic long-range FM order in pristine Cr$_2$Ge$_2$Te$_6$ atomic layers has been reported that the Curie temperature ($T_C$) can be controlled by the very small magnetic fields (smaller than 0.3 T).\cite{add3} Furthermore, the feasibility of exfoliation in experiment is confirmed by evaluating the cleavage energy to be very close to graphite.\cite{22} $T_C$ of monolayer CrSnTe$_3$ is determined as 170 K, higher than monolayer CrSiTe$_3$ and CrGeTe$_3$. This is attribute to the increased super-exchange coupling between the magnetic Cr atoms by enhanced ionicity of the Sn-Te bond.\cite{23}

Due to the 2D magnetism in above ABX$_3$-class materials, we use elemental substitution, one of the commonest experimental methods, to improve the physical properties. In this work, we investigate electronic and magnetic structures of monolayer CrGeS$_3$ based on first-principles calculations. The in-plane atoms of CrGeS$_3$ combine with each other by intense covalent bond while the weak van der Waals bonding exists between layers. Additionally, the exfoliation energy is similar to graphene, therefore, we harbor the idea that CrGeS$_3$ can be exfoliated into monolayer successfully. Our results demonstrate monolayer CrGeS$_3$ is dynamically and mechanically stable and possesses intrinsic FM. The electronic structures show that monolayer CrGeS$_3$ is FM semiconductor with global bandgap of 0.34 eV (2.21 eV by HSE06 calculations) and possesses a bandgap of 0.70 eV and 0.77 eV in spin-up and spin-down states, respectively. The magnetic properties result from competition between FM hybridized S $p$ holes and AFM itinerant holes in S $p$ orbits. Furthermore, electronic exchange interaction and orbital hybridizations in monolayer CrGeS$_3$ crystal lead to the $T_C$ of 161 K in the mean-field theory (MFT). The present results appear promising platform for studying fundamental spin behaviours and promote applications in ultra-compact spintronics.
\section{computational details}
Electronic structure and magnetic properties were carried out using density functional theory (DFT) by the projector augment wave (PAW) method as implemented in the Vienna ab initio simulation package (VASP).\cite{24, 25} Perdew-Burke-Ernzerhof (PBE) was applied to handle with Exchange-correlation function.\cite{26} The accurate global band gap was calculated using the Heyd-Scuseria-Ernzerhof (HSE06) functional including 25\% non-local Hartree-Fock exchange.\cite{HSE}. Cut-off energy for plane wave basis was set up to 450 eV. The number of k points were set as $11\times11\times1$ for the monolayer structure.  Note that the cutoff energy and the number of k points were all tested assuring that geometry structure gave entire relaxation and sites of atoms reached convergence. The same test was applied in the choice thickness of vacuum slab as 15 \AA\ to separate correlation between adjacent layers. A convergence criterion of $10^{-6}$ for total energy of electronic consistence loop was employed, and 0.01 eV/ \AA\ for Hellmann-Feynman force components in the ions relaxation loop. The partial occupancies were set for each orbital with Gaussian smearing and the width of the smearing was set up to 0.05 eV. A supercell size of $3\times3\times1$ primitive unit cells is built to calculated phonon band spectrum by using the density functional perturbation theory (DFPT)~\cite{Baroni2001}.

\section{result and discussion}

\begin{figure}[htbp]
\centering
\includegraphics[scale=0.4]{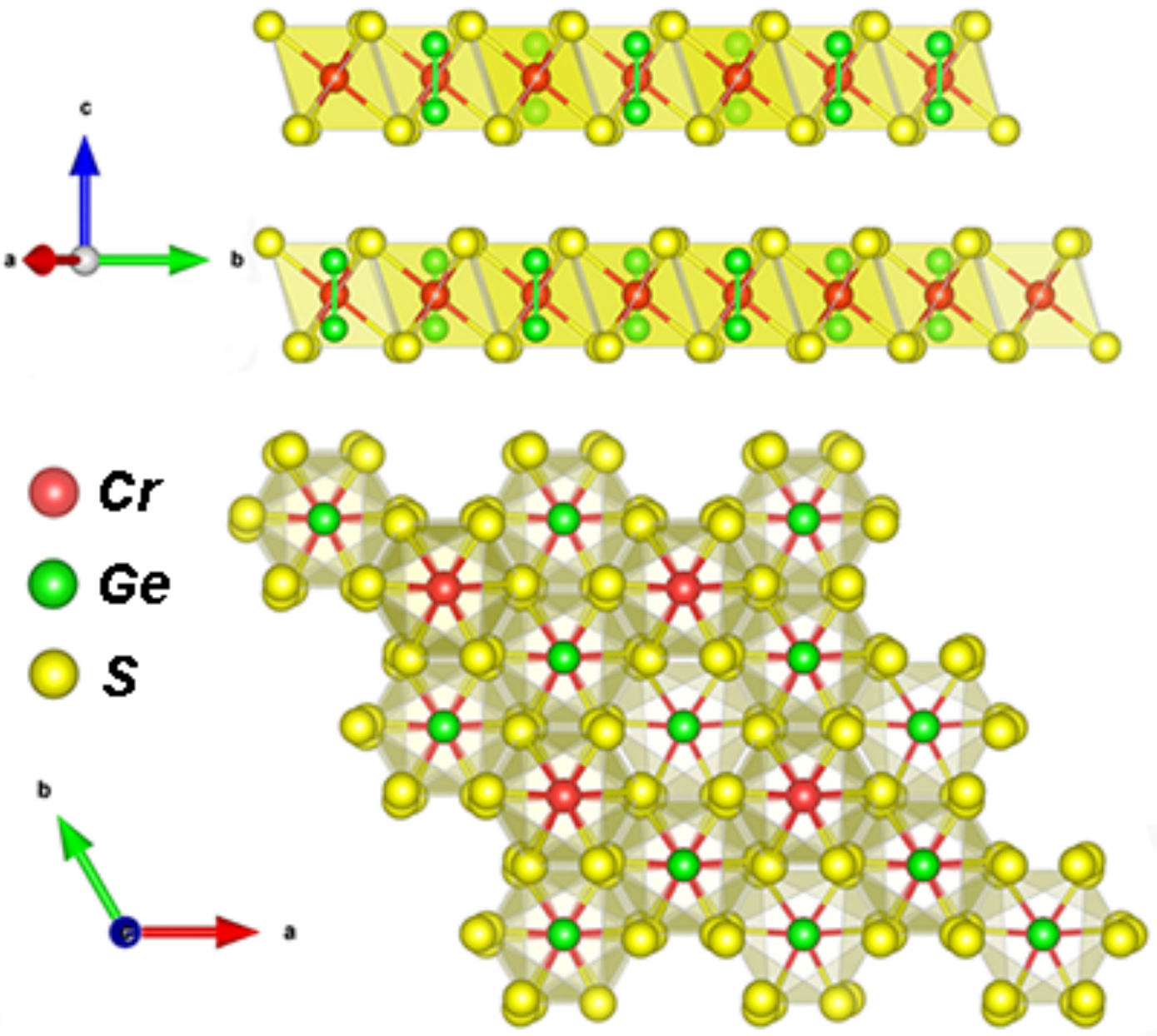}
        \caption{\label{fig1}Top (a) and side (b) views of monolayer CrGeS$_3$ structure. The Cr, S and Ge atoms are displayed as red, yellow and green spheres respectively.}
       \end{figure}

CrGeS$_3$ is a layered material crystallized in space group R $\overline{3}$ (No. 148) and formed by stacks of S-(Ge,Cr)-S sandwich layers with lattice parameters of a=b=6.05 \AA. Due to the smaller radius of S atoms than Te atoms, the lattice parameters of $ab$ plane exhibit a shrinkage compared with a=6.86 \AA\ of CrGeTe$_3$.\cite{add2} As shown in fig.\ref{fig1}, the monolayer hexagonal honeycomb structure is composed of a Ge$_2$S$_6$ with two Cr ions inserted between two-layers S planes. The monolayer unit cell is composed of two Cr$^{2+}$ ions and one [Ge$_2$S$_6$]$^{4-}$. Each Ge atom possesses three neighbouring S atoms forming a tetrahedron and two of Ge-centered tetrahedrons form a dumbbell-like Ge dimer in [Ge$_2$S$_6$]$^{4-}$ bipyramid. The innerlayer atoms are combined by strong covalent bond, while the weak van der Waals interaction exists between layers. Bonds length of Ge-Ge, Cr-S and Ge-S are 2.34 \AA, 2.45 \AA\ and 2.23 \AA, respectively.

\begin{figure}[htbp]
\centering
\includegraphics[scale=0.5]{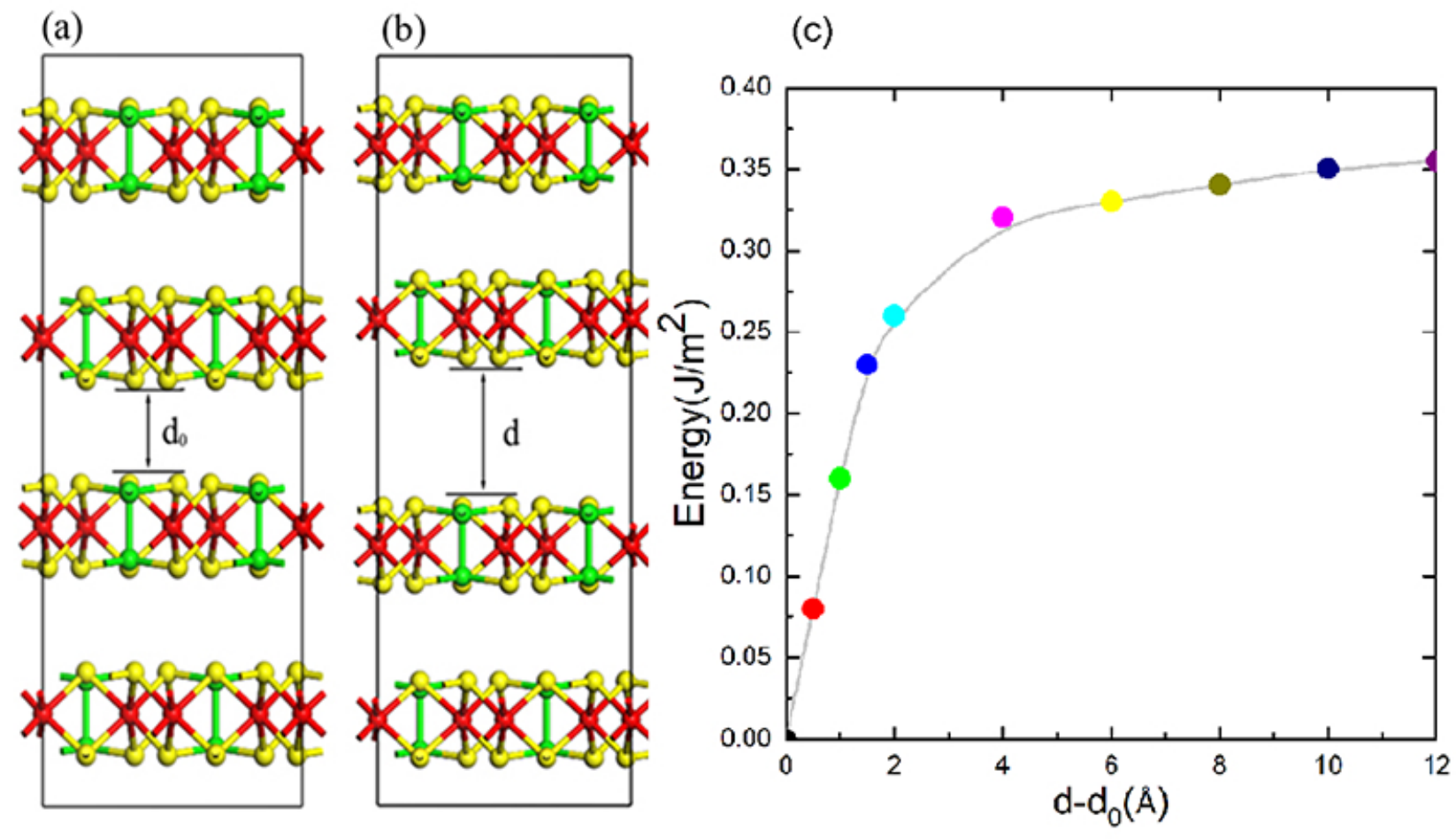}
       \caption{\label{fig2}(a)Crystal structure of bulk CrGeS$_3$ in ground state. $d_0$ is the distance between layers of optimized bulk CrGeS$_3$ crystal. (b) Exfoliating simulation procedure with an increasing value of $d$, and the $d$ is distance of two divided parts. (c) Exfoliation energy as a functional of the increasing value of $d$.}
     \end{figure}

Exfoliation energy is the energy required to cut bulk materials into two halves and is estimated by gradually increasing the distance of two parts (fig.\ref{fig2}). The variation of exfoliation energy with the distance $d$ [fig.\ref{fig2}(c)] reveals that the energy required to break out-of-plane van der Waals binding exhibits an abrupt growing below d=4 \AA, while there is a sharp decrease of interaction between two parts. Then the exfoliation energy achieves convergence at an ideal value of 0.35 J/$m^2$. This value closes to 0.36 J/$m^2$ in graphene\cite{28} and ensures the experimental synthesis of monolayer CrGeS$_3$.

\begin{figure}[htbp]
\centering
\includegraphics[scale=3.3]{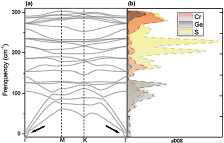}
       \caption{\label{fig3}(a) Phonon band structure of monolayer CrGeS$_3$, two small pockets of imaginary frequency are labelled as black arrow. (b) Corresponding project phonon density of states (pDOS).}
     \end{figure}

In order to check the dynamic and mechanical stability of the monolayer structure, we calculate phonon spectrum and projected density of states of phonon, as shown in fig.\ref{fig3}. Two negligible small pockets of instability occur near $\Gamma$ point labelled as black arrow [fig.\ref{fig3}(a)]. The pockets are extremely sensitive to computational details. Also, there is the altogether absence of these imaginary frequency in some cases with respect to our calculations. Note that it is difficult to achieve numerical convergence for the flexural phonon branch, which is considered as a common issue in first-principles calculations for two-dimensional materials.\cite{pho1, pho2, pho3, pho4} Therefore, the phonon spectrum suggests that monolayer CrGeS$_3$ is dynamically stable.

The energetic stability of monolayer CrGeS$_3$ can be evaluated by formation energy E$_f$ as
\begin{eqnarray}
E_f =\frac{E\left(CrGeS_3\right)-E\left(Cr\right)-E\left(Ge\right)-3\cdot E\left(S\right)}{5}
\end{eqnarray}
where $E$ is the corresponding normalized energy for different structures. The negative $E_f$ of -0.61 eV indicates monolayer CrGeS$_3$ is energetically stable. For a further detecting mechanical stability, we also calculate the elastic constants $C_{11}$ and $C_{12}$ of monolayer CrGeS$_3$. The two independent parameters are given as,
        \begin{eqnarray}C_{11} = \frac{1}{A_0} \cdot \frac{\partial^2 E_{tot}}{\partial \varepsilon_{11}^2}\hspace{1em}and\hspace{1em}C_{12} = \frac{1}{A_0} \cdot \frac{\partial^2 E_{tot}}{\partial \varepsilon_{11} \partial \varepsilon_{12}}
       \end{eqnarray}
where $E_{tot}$ is total energy of an unit cell and $A_0$ is the equilibrium area of monolayer CrGeS$_3$. The calculated $C_{11}$ and $C_{12}$ are 64 N/m and 17 N/m, respectively. It is evident that our results satisfy the Born criterion of stability ($C_{11}-C_{12}\ge0$)\cite{29}.

\begin{figure}[htbp]
  \centering
  \includegraphics[scale=0.75]{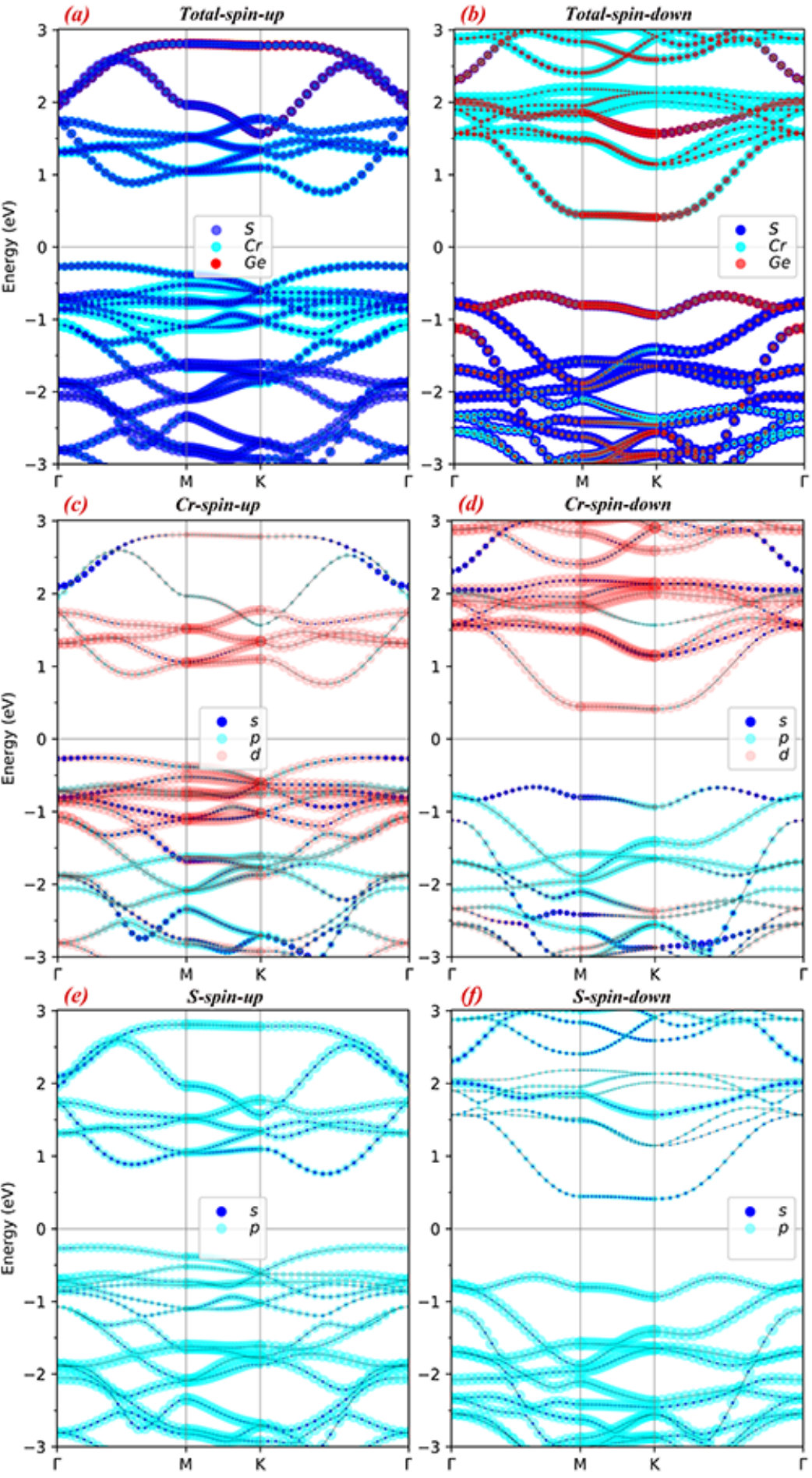}
        \caption{\label{fig4}Electronic band structures of single-layer CrGeS$_3$ obtained from first-principles calculations using density-functional theory (DFT). (a) and (b) the electronic band structures of spin-up and spin-down are denoted as blue, cyan and red respectively. (c) and (d) the projected electronic band structures for spin-up and spin-down states of Cr atoms are denoted by blue, cyan and red for $s$, $p$, $d$ orbits respectively. (e) and (f) the corresponding projected electric band structures of S atoms.}
        \end{figure}

For exploring into the electronic properties of monolayer CrGeS$_{3}$, we plot the spin polarized band structure diagrams as shown in fig.\ref{fig4}. The spin-down components exhibit similar mechanism of electronic interaction of conduction bands and valence bands with spin-up state, so the spin-up components are discussed here mainly [fig.\ref{fig4}(a), (c) and (e)]. The spin-up band structure possesses the bandgap of 0.70 eV. The conduction band originates from significant hybridization between empty $dp$ antibonding bands of Cr $d$-S $p$ orbits at $\thicksim$1 eV above the Fermi level when valence band is the fully occupied $dp$-hybridized bonding states of Cr $d$-S $p$ at about -0.5 eV under the Fermi level. The highly localized antibonding bands of Cr atoms are mainly composed of unoccupied, polarized $d$ orbits. In comparison, an effective exchange interaction of the bonding-states Cr atoms is generated by the $s$-$d$ hybridization. Additionally, the occupied Ge $sp^3$-S $p$ hybridization in spin-down channel [fig.\ref{fig4}(b), (d) and (f)] results from the tetrahedral-coordinated Ge atom surrounded by three S atoms in [Ge$_2$S$_6$]$^{4-}$, so the Ge $sp^3$ hybrid orbits lead to bonding states of Ge $sp^3$-S $p$.

\begin{figure}[htbp]
  \centering
  \includegraphics[scale=1.1]{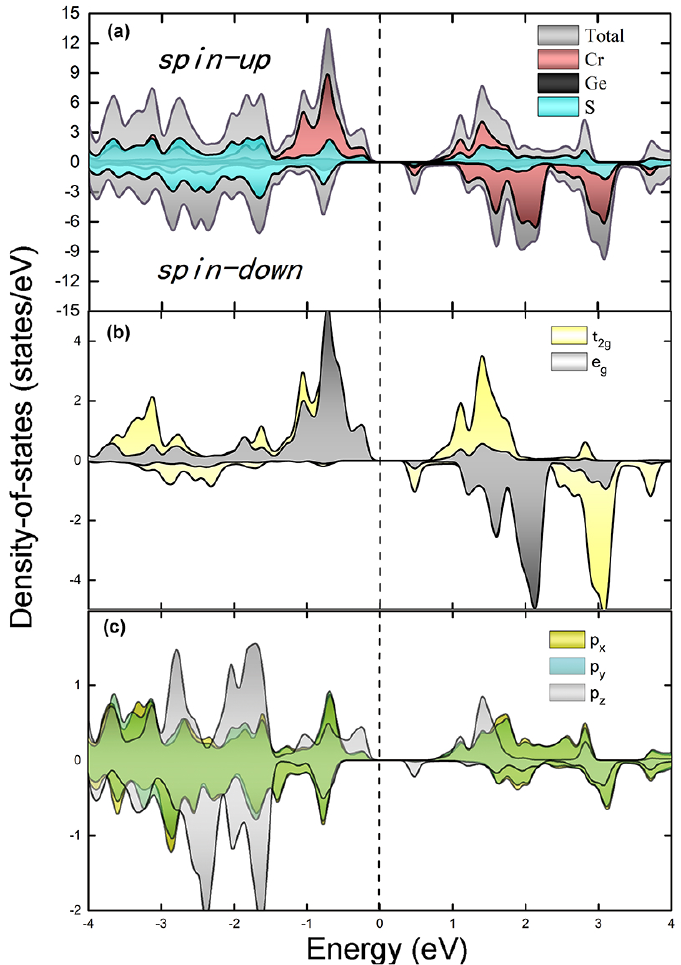}
        \caption{\label{fig5}(a) total density of state (DOS) of monolayer CrGeS$_3$, (b) project density of state (pDOS) of Cr and (c) S. In the pDOS plot, the upper panel represents for the spin-up components and the lower panel for the spin-down components. The Fermi level is set to zero.}
        \end{figure}

To help a further understanding of the electronic structure of monolayer CrGeS$_3$, we present the project density of states based on PBE functional as shown in fig.\ref{fig5}. Figure 5 (a) ensures the global bandgap of 0.34 eV and a more accurate value is investigated as 2.21 eV based on HSE06 functional. The CrGeS$_3$ crystal field splits the $3d$ orbits of Cr atoms into upper $e_g$ ($d_{x^2-y^2}$, $d_{z^2}$) and lower $t_{2g}$ ($d_{xy}$, $d_{zx}$, $d_{yz}$) states, as shown in fig.\ref{fig5}(b). The prominent feature of the electronic structure is the fully occupied Cr $e_g$ bands at about -0.5 eV under the Fermi level and the empty spin-down $e_g$ orbits of Cr atoms localizing at about 2.1 eV above the Fermi level, indicating a spin split (about 0.42 eV) between the localized Cr $e_g$ states. The bonding $dp\sigma$ hybridized states of Cr $e_g$-S $p$ and the unoccupied Cr $t_{2g}$-Ge $p$ hybridization results in valence band maximum and conduction band minimum. Apart from this, each $e_g$ orbits of Cr atom hybridizes significantly with S $p$ states, forming a bonging and antibonding pair bonds of Cr $e_g$-S $p$. But the weaker $dp\pi$ bonding gives rise to the localized Cr $t_{2g}$ states, which is agreed with former research on ABX$_3$ type transition metal tri-chalcogenides.\cite{add1}

Now we estimate the magnetic properties by comparing FM and AFM phase of monolayer CrGeS$_3$. The results show the AFM state possesses a higher energy about 50 meV than FM state. The FM coupling is contributed by the hybridized $p$ holes of S atoms. And the $p$ holes are originated from strong $dp\sigma$-hybridization between Cr $d$ and $p$ orbits of S atoms. Apart from the regular super-exchange interactions mentioned above, the itinerant holes in $p$ orbits of S atoms can also be coupled with the neighboring Cr spins antiferromagnetically, leading to a mediation of FM-states moments of Cr atoms. Thus, the local magnetic moment of one Cr atom is 3.27 $\mu_{B}$ with an extra 0.27 $\mu_{B}$ originating from the bonding states of Cr $e_g$-S $p$. Actually, this interaction of Cr $d$ orbits and S $p$ leads to the S $p$ holes with an opposite spin polarization of -0.11 $\mu_{B}$ each S atom according to our results. Then the magnetic moment of a Cr atom remains 3 $\mu_{B}$.

For magnetic materials, Curie temperature $T_C$ is a crucial parameter determining a temperature at which intrinsic magnetic intensity decrease to 0 and FM phase transforms into nonmagnetic phase.\cite{30, 32, 33} We main focus on predicting $T_C$ in accordance with MFT\cite{34} as
          \begin{eqnarray}T_C=\frac{2zJS\left(S+1\right)}{3k_B}
          \end{eqnarray}

where $z=3$ is the number of nearest-neighboring Cr atoms of a Cr atom in monolayer CrGeS$_3$, $J$ is the exchange integral, $S=3/2$ is the spin of each Cr atom, and $k_B$ is the Boltzmann constant. The value of $S$ is consistent with the magnetic moment of 3 $\mu_{B}$ for each Cr atom. We induce another term $E_{ex}$ as exchange energy between the two Cr atoms in unit cell, defined as
           \begin{eqnarray}E_{ex}=-2zJS^2
           \end{eqnarray}
The total energies of monolayer CrGeS$_3$ primitive cell with the FM and AFM configurations $E_{FM}$ and $E_{AFM}$ can be calculated as well.\cite{35} Accordingly, the difference between $E_{FM}$ and $E_{AFM}$ gives $2E_{ex}$. As a result, we can obtain the exchange integral $J$ with the PBE functional as 1.85 meV. We further obtain $T_C$ of 161 K in monolayer CrGeS$_3$. For comparison, the $T_C$ of CrGeTe$_3$ with a similar structure via Monte Carlo (MC) simulation is 61 K\cite{36}. Note that the MFT generally overestimates the $T_C$\cite{37} compared with MC simulations and we just provide a prediction of monolayer CrGeS$_3$. The properties suggest that monolayer CrGeS$_3$ have potential applications in spin-polarized carrier injection, detection, ultra-compact spintronics, and also support a platform for studying electronic mechanism in 2D materials.\cite{22}
\section{Conclusion}
In summary, we have investigated the stability, electronic and magnetic properties of monolayer CrGeS$_3$ based on first-principles calculations. It is found that monolayer CrGeS$_3$ is dynamically and mechanically stable and could be successfully synthesized due to the similar exfoliation energy to graphene. After that, honeycomb structural monolayer CrGeS$_3$ exhibits FM ground state with magnetic moment of 3 $\mu_{B}$ per Cr atoms. The reason of this magnetism is the competition between $dp\sigma$ bonding states of Cr $e_g$-S $p$ intercoupling ferromagnetically and opposite spin polarization from S $p$ holes. The indirect bandgaps are 0.34 eV (2.21 eV based on HSE06 functional), 0.70 eV and 0.77 eV in global, spin-up and spin-down states respectively, which result from the interaction between bonding $dp\sigma$ hybridized states of Cr $e_g$-S $p$ and unoccupied bonds of Cr $t_{2g}$-Ge $p$. In addition, we roughly estimate the $T_C$ of magnetic transition as 161 K by MFT. Our results illustrate that monolayer CrGeS$_3$ crystals will possess enormous applications in nanoscale spintronics. We also expect our theoretical study will inspire further experimental studies.
\begin{acknowledgments}
This work was supported by the NSFC (Grants No.11874273), the Specialized Research Fund for the Doctoral Program of Higher Education of China (Grant No.2018M631760), the Project of Heibei Educational Department, China (No.ZD2018015), the Advanced Postdoctoral Programs of Hebei Province (No.B2017003004).
\end{acknowledgments}

\end{document}